\documentclass{PoS}
\usepackage{amsmath}
\usepackage{calc}
\usepackage{graphicx}
\usepackage{color}
\graphicspath{{./figures/}{./}}
\newcommand{\dint}{{\rm{d}}}
\newcommand{\Dint}{{\rm{D}}}
\newcommand{\ri}{{\rm{i}}}

\title{Calculating the static gravitational two-body potential to fifth post-Newtonian order with Feynman diagrams}

\ShortTitle{Static gravitational two-body potential to fifth post-Newtonian order}

\author{Stefano Foffa\\%
        D\'epartement de Physique Th\'eorique and Centre for Astroparticle
        Physics,\\
        Universit\'e de Gen\`eve,
        CH-1211 Geneva,
        Switzerland\\
        E-mail: \email{stefano.foffa@unige.ch}
       }

\author{Pierpaolo Mastrolia\\%
        Dipartimento di Fisica ed Astronomia,
        Universit\`a di Padova,
        Via Marzolo 8,
        35131 Padova,
        Italy\\
        INFN, Sezione di Padova, Via Marzolo 8, 35131 Padova, Italy \\
        E-mail: \email{pierpaolo.mastrolia@pd.infn.it}
       }

\author{Riccardo Sturani\\%
        International Institute of Physics (IIP),\\ 
        Universidade Federal do Rio Grande do Norte (UFRN) CP 1613,\\
        59078-970 Natal-RN,
        Brazil\\
        E-mail: \email{riccardo@iip.ufrn.br}
       }

\author{\speaker{Christian Sturm}\\
        Universit{\"a}t W{\"u}rzburg, 
        Institut f{\"u}r Theoretische Physik und Astrophysik,\\
        Emil-Hilb-Weg 22, 
        D-97074 W{\"u}rzburg,
        Germany\\
        E-mail: \email{christian.sturm@physik.uni-wuerzburg.de}%
       }

\author{William J. Torres Bobadilla\\%
        Instituto de F\'{\i}sica Corpuscular,
        Universitat de Val\`{e}ncia -- Consejo Superior de
        Investigaciones Cient\'{\i}ficas, 
        Parc Cient\'{\i}fic,
        E-46980 Paterna, Valencia,
        Spain\\
        E-mail: \email{william.torres@ific.uv.es}
       }

\abstract{%
We discuss the first-time calculation of the static gravitational two-body
potential up to fifth post-Newtonian(PN) order. The results are achieved
through a manifest factorization property of the odd PN diagrams. The
factorization property is illustrated also at first and third PN order.
}
 \FullConference{14th International Symposium on Radiative Corrections (RADCOR2019)\\ 
 9-13 September 2019\\
 		Palais des Papes, Avignon, France}

\begin{document}
\section{Introduction\label{sec:Intro}}
The classic Newton potential of two gravitationally interacting massive
bodies receives post-New\-tonian(PN) corrections due to effects of
general relativity(GR). They can be calculated systematically in
perturbation theory in the non-relativistic limit for weak curvature and
small velocities. The expansion is performed in virial-related
quantities like the relative squared velocity $v^2\sim G_N\*m/r$ and the
compactness $R_S/r\sim G_N\*m/r$, where $G_N$ is the Newton constant,
$m$ is the mass, $r$ the relative distance of the binary components and
$R_S$ is the Schwarzschild radius.
A given $n$-th PN order has a manifest power counting in terms of powers
$\ell$ of the Newton constant $G_N$ and powers $k$ of the velocities $v$
squared.  The $n$-th PN order is then given by $n=k+\ell-1$.

The first PN order is known already since long and was determined by
Einstein, Infeld and Hoffmann~\cite{Einstein:1938yz}.  
It helped in the understanding of phenomena which are observable within
our own solar system and which arise due to effects of GR, like for
example, it contributes to explain the perihelion precession of mercury.

The direct observation of gravitational waves emitted by a coalescing
binary system through the LIGO and Virgo
collaborations~\cite{Abbott:2016blz} was a tremendous success and probe
of GR. The PN corrections are here important for the
construction of wave form templates which are used in the
LIGO/Virgo~\cite{TheLIGOScientific:2014jea,TheVirgo:2014hva} data
analysis pipeline~\cite{Taracchini:2012ig,Schmidt:2014iyl} for the
detection of the gravitational waves.
The second and third PN order has been calculated 
in refs.~\cite{Damour:1982wm,Damour:1985mt}
and in refs.~\cite{Damour:2001bu,Blanchet:2003gy,Itoh:2003fy},
respectively. 
The fourth PN order was first determined in 
\cite{Damour:2014jta,Damour:2015isa,Damour:2016abl}  
and confirmed by  
\cite{Bernard:2015njp,Bernard:2016wrg,Bernard:2017bvn,Marchand:2017pir,Bernard:2017ktp}
and \cite{Foffa:2012rn,Foffa:2016rgu,Foffa:2019rdf,Foffa:2019yfl}.
Future observatories, like the Einstein Telescope~\cite{Punturo:2010zz}
and LISA~\cite{Audley:2017drz} are expected to gain at least an order of
magnitude in sensitivity with respect to current observatories. As a
result of this also an increased precision of the theory description is
desirable~\cite{Lindblom:2008cm,Antonelli:2019ytb}.

In general there are different approaches to solve the gravitational
two-body problem, where in the following we will focus solely on the
effective field theory~(EFT)
approach~\cite{Goldberger:2004jt,Goldberger:2007hy,Foffa:2013qca,Rothstein:2014sra,Porto:2016pyg,Levi:2018nxp}. In
the EFT approach the problem of computing PN corrections to the
gravitational two-body potential can be traced back to the calculation
of Feynman diagrams. In particular the rich methodology commonly used in
particle physics for the determination of loop integrals is in this
approach directly applicable in order to accomplish such
calculations. In the following we focus on the determination of the
conservative sector of the gravitational two-body potential in the
static limit up to fifth PN order.
\section{The effective action}
Following the lines of refs.~\cite{Foffa:2011ub,Foffa:2012rn}, the
action $S$ which describes the gravitational interaction can 
be decomposed into two contributions $S=S_{\rm{pp}}+S_{\rm{bulk}}$. 
The world-line point-particle action $S_{\rm{pp}}$ is representing the
two binary components with masses $m_1$ and $m_2$. They are considered
as spinless point masses. We also neglect tidal effects. The
point-particle action reads:
\begin{equation}
S_{\rm{pp}}=-\sum_{i=1,2}m_i\int\dint\tau_i
        =-\sum_{i=1,2}m_i\int\sqrt{-g_{\mu\nu}(x_i)\dint x_i^\mu\dint x_i^\nu}.
\end{equation}
The bulk action $S_{\rm{bulk}}$ consists out of
the Einstein-Hilbert action plus a gauge fixing term~\cite{lrr-2002-3,Bernard:2015njp}:
\begin{equation}
S_{\rm{bulk}}=2\Lambda^2\int \dint^{d+1}x\sqrt{-g}
           \left[R(g)-\frac{1}{2}\Gamma^{\mu}\Gamma_{\mu}\right],
\end{equation}
where the harmonic gauge condition has been adopted and $\Gamma^{\mu}$
is given through the Christoffel symbol $\Gamma^{\mu}_{\rho\sigma}$ in
the equation
$\Gamma^{\mu}=g^{\rho\sigma}\Gamma^{\mu}_{\rho\sigma}$. Furthermore,
$\Lambda^{-2}=32\pi\,G_N L^{d-3}$, where $d$ is the spatial dimension
and $L$ an arbitrary length scale which takes care about the proper mass
dimension in dimensional regularization. It vanishes in physical
observables in the limit $d\to3$. For the metric tensor we use the
Kaluza-Klein parametrization~\cite{Kol:2007bc,Kol:2007rx}
\begin{equation}
  g_{\mu\nu}=e^{2\phi/\Lambda}
  \begin{pmatrix}
    -1        \quad  & A_j/\Lambda\\
    A_i/\Lambda\quad & e^{-c_d\phi/\Lambda}\gamma_{ij}-A_iA_j/\Lambda^2
  \end{pmatrix},
\end{equation}
where its degrees of freedom are parametrized by three fields: a scalar
field $\phi$, a vector field $A_i$ and a symmetric tensor field
$\sigma_{ij}$.  The symbols $c_d$ and $\gamma_{ij}$ are given by
$c_d=2(d-1)/(d-2)$ and $\gamma_{ij}=\delta_{ij}+\sigma_{ij}/\Lambda$,
where the indices $i$, $j$ run over all $d$ spacial dimensions. It turns
out that in the static limit the vector fields $A_i$ do not contribute
to our calculation. The effective action is obtained by integrating out
the remaining gravity fields
\begin{equation}
\exp[\ri S_{\rm eff}]=\int\Dint\phi\Dint\sigma_{ij} \exp\left[\ri\left(S_{\rm{pp}}+S_{\rm{bulk}}\right)\right],
\end{equation}
which is perturbatively expanded.\\
\begin{figure}[!h]
\begin{center}
 \raisebox{0.6cm}{
 \begin{minipage}{2.9cm}
 \hspace*{0.8cm}\rotatebox[origin=c]{180}{$\hookleftarrow$}{\footnotesize{\mbox{$\sim 
       m /(k!\Lambda^{k})$}}}\\
\includegraphics[width=2.1cm]{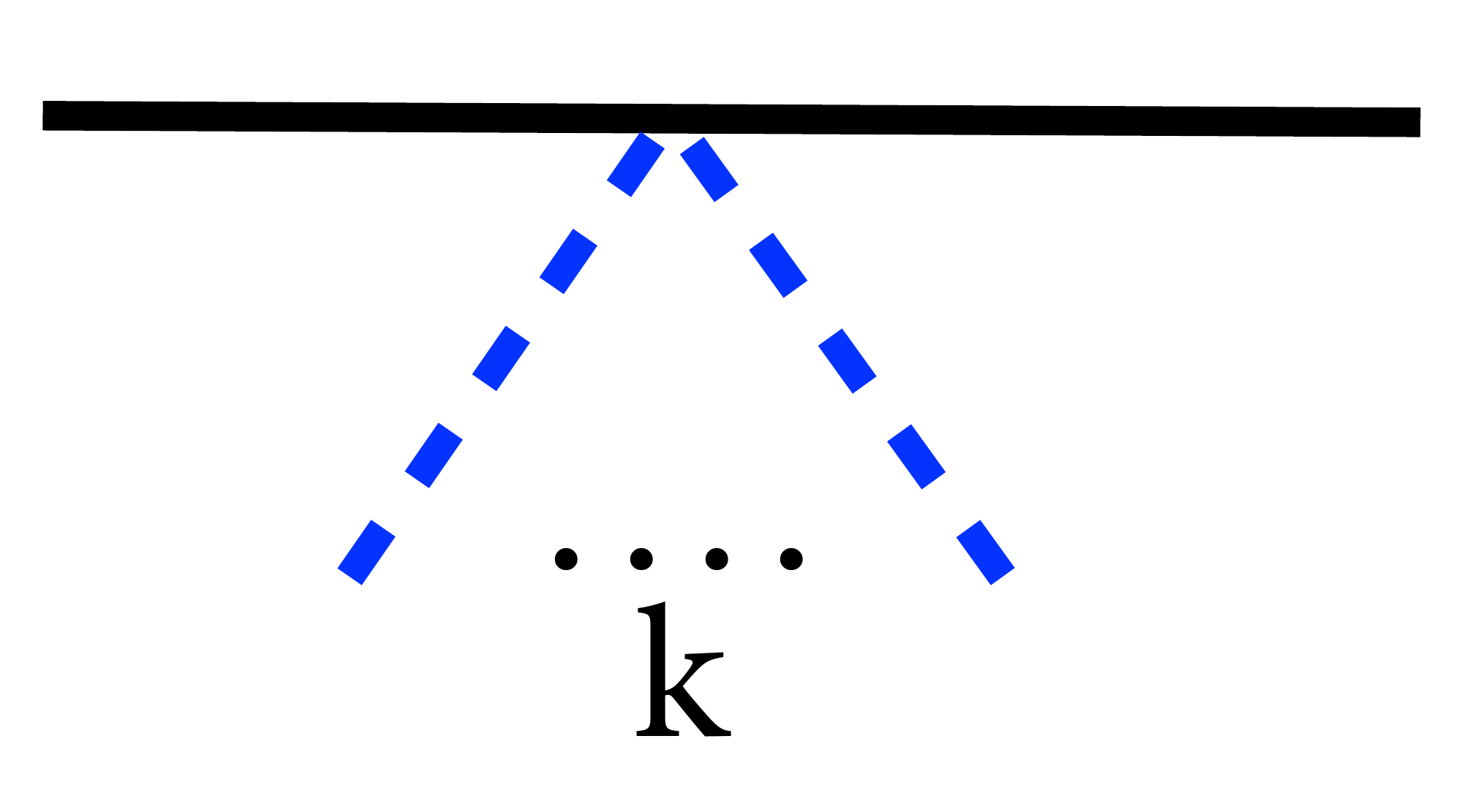}
 \end{minipage}
 }
 \hspace{0.5cm}
\includegraphics[width=2.7cm]{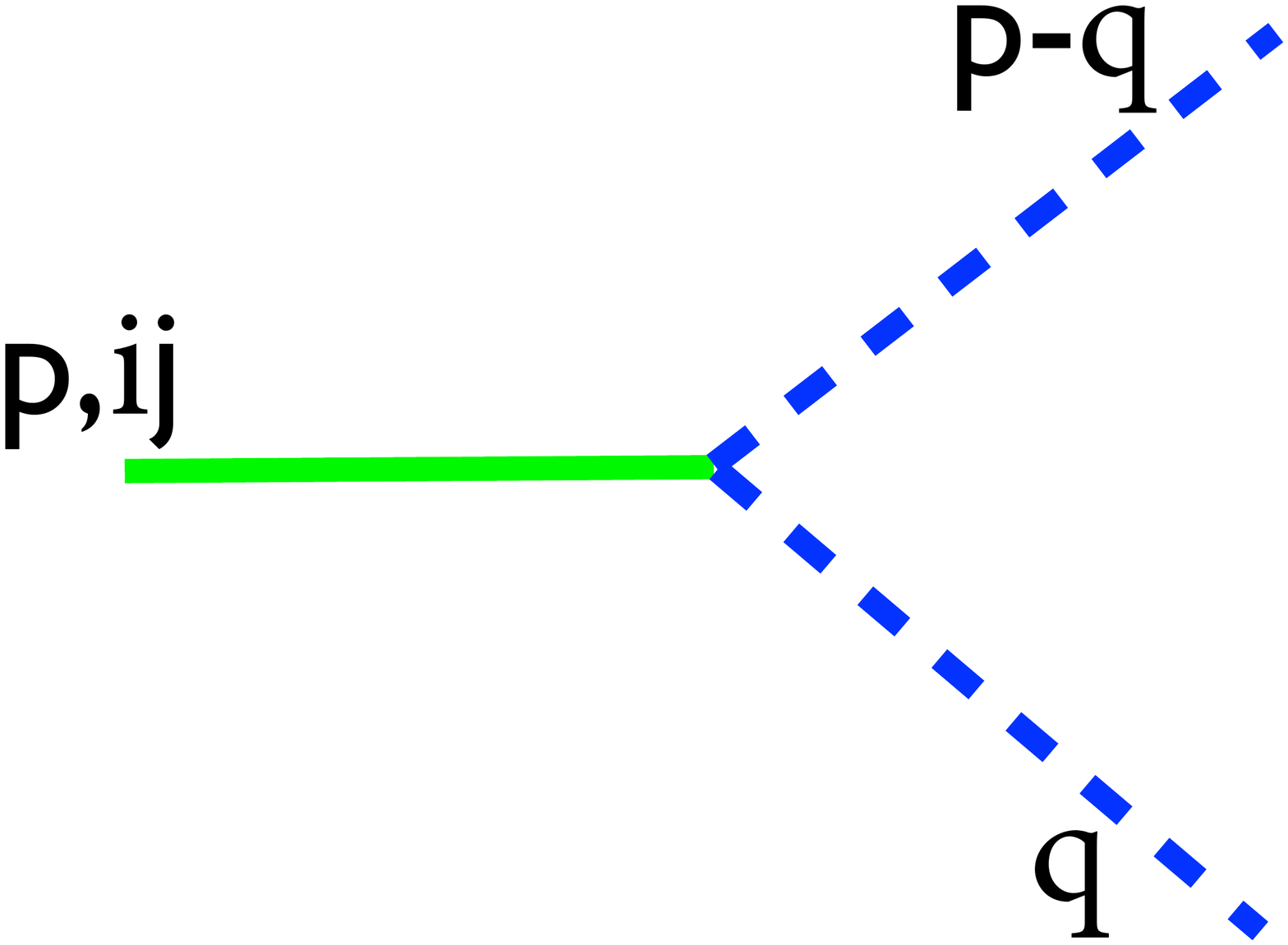}
\hspace{0.5cm}
\includegraphics[width=2.7cm]{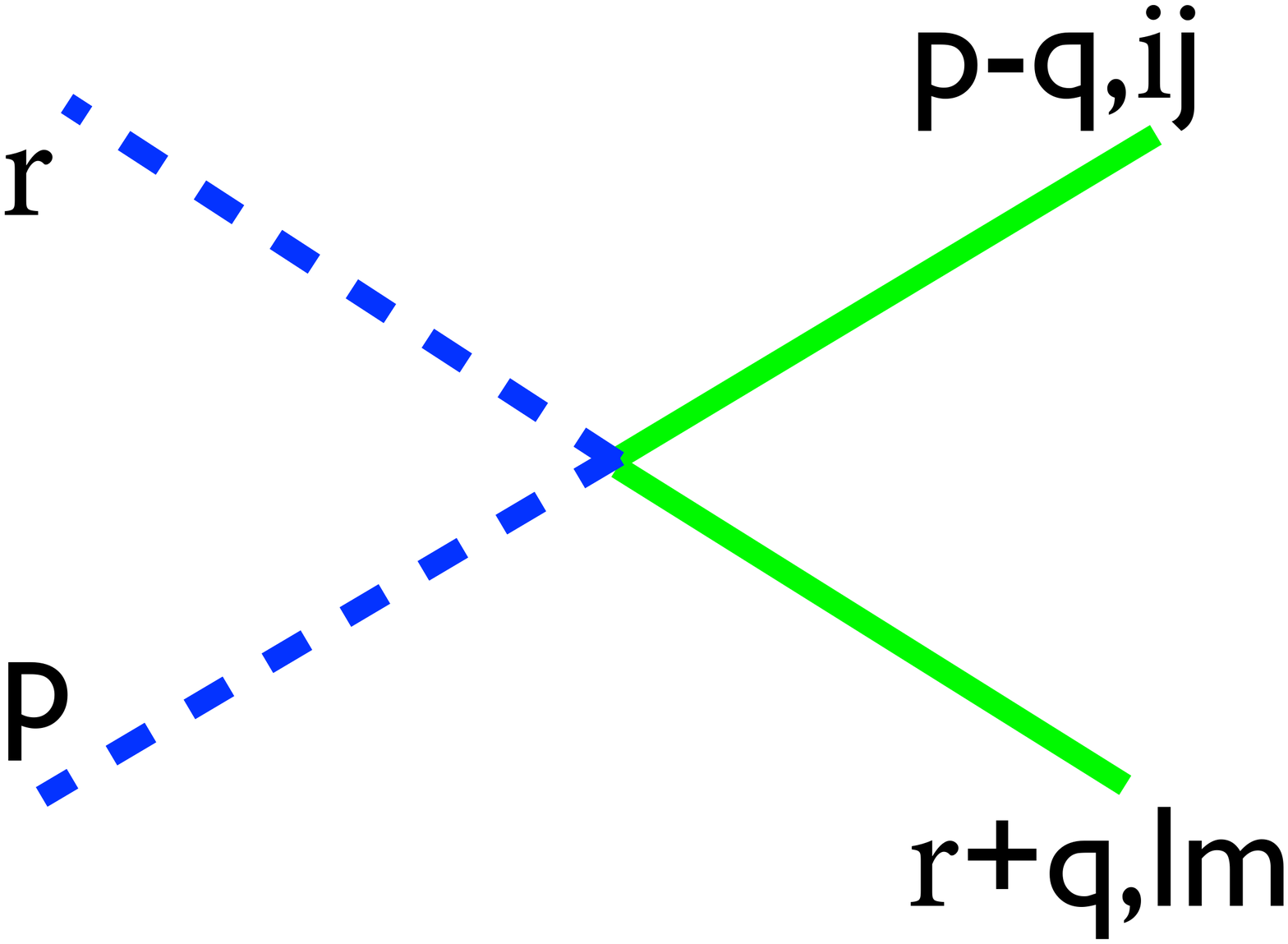}
\hspace{0.5cm}
\includegraphics[width=2.7cm]{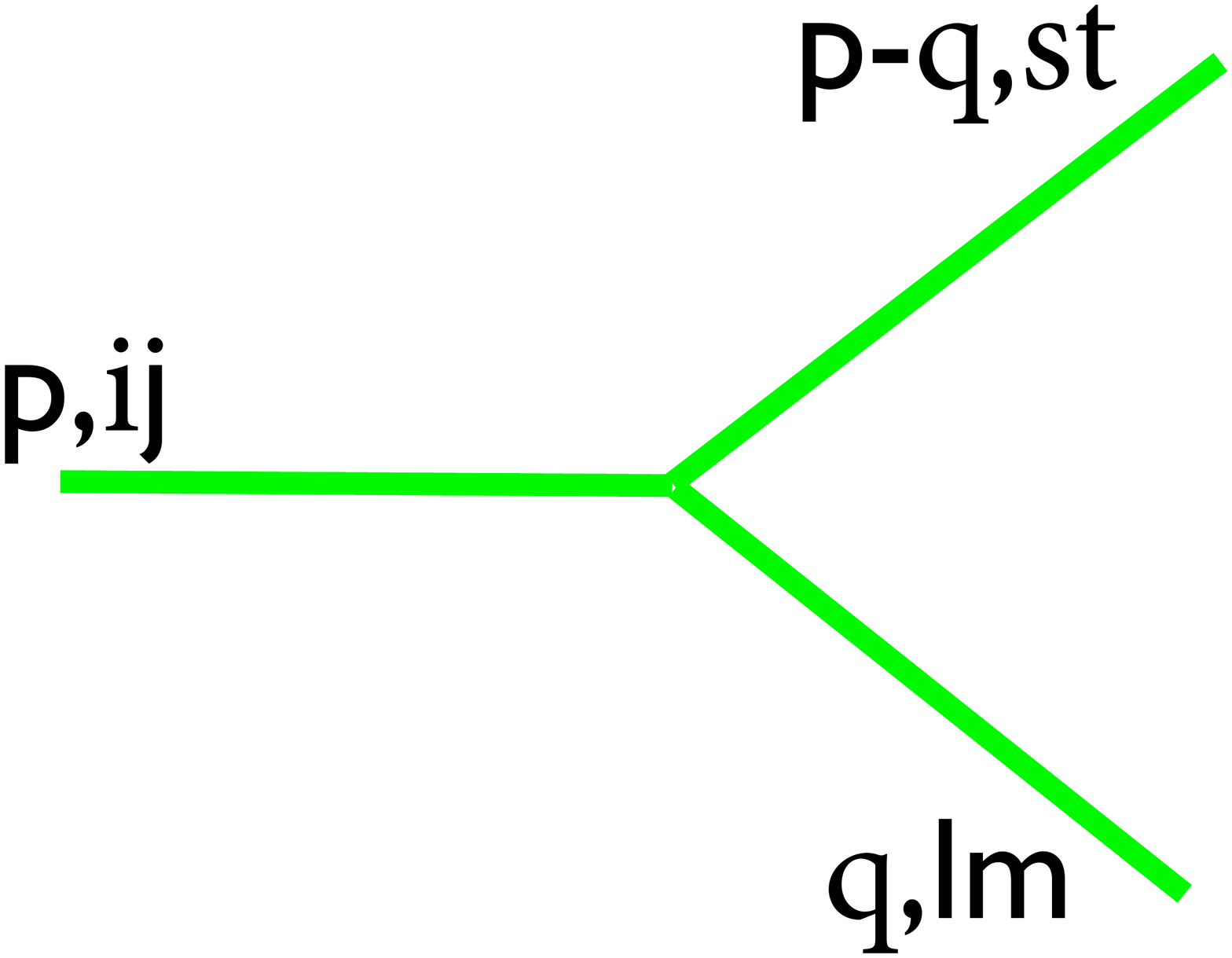}
\end{center}
\vspace*{-0.4cm}
\caption{The vertices which contribute to the calculation of the
  static two-body potential up to fifth PN order are shown.
 The first diagram shows the matter-$\phi^k$ vertex, where the solid
 black lines stand for the non-dynamical external sources.  
  In the three bulk vertices the solid green lines
  correspond to tensor fields, whereas the blue dashed lines correspond
  to scalar fields. The external momenta are $p$, $q$, $r$ and the
  indices of the tensor fields are given by $i$, $j$, $l$, $m$, $s$, $t$.
  \label{fig:vertices}} 
\end{figure}

\begin{figure}
\begin{center}  
  \includegraphics[width=1.3cm]{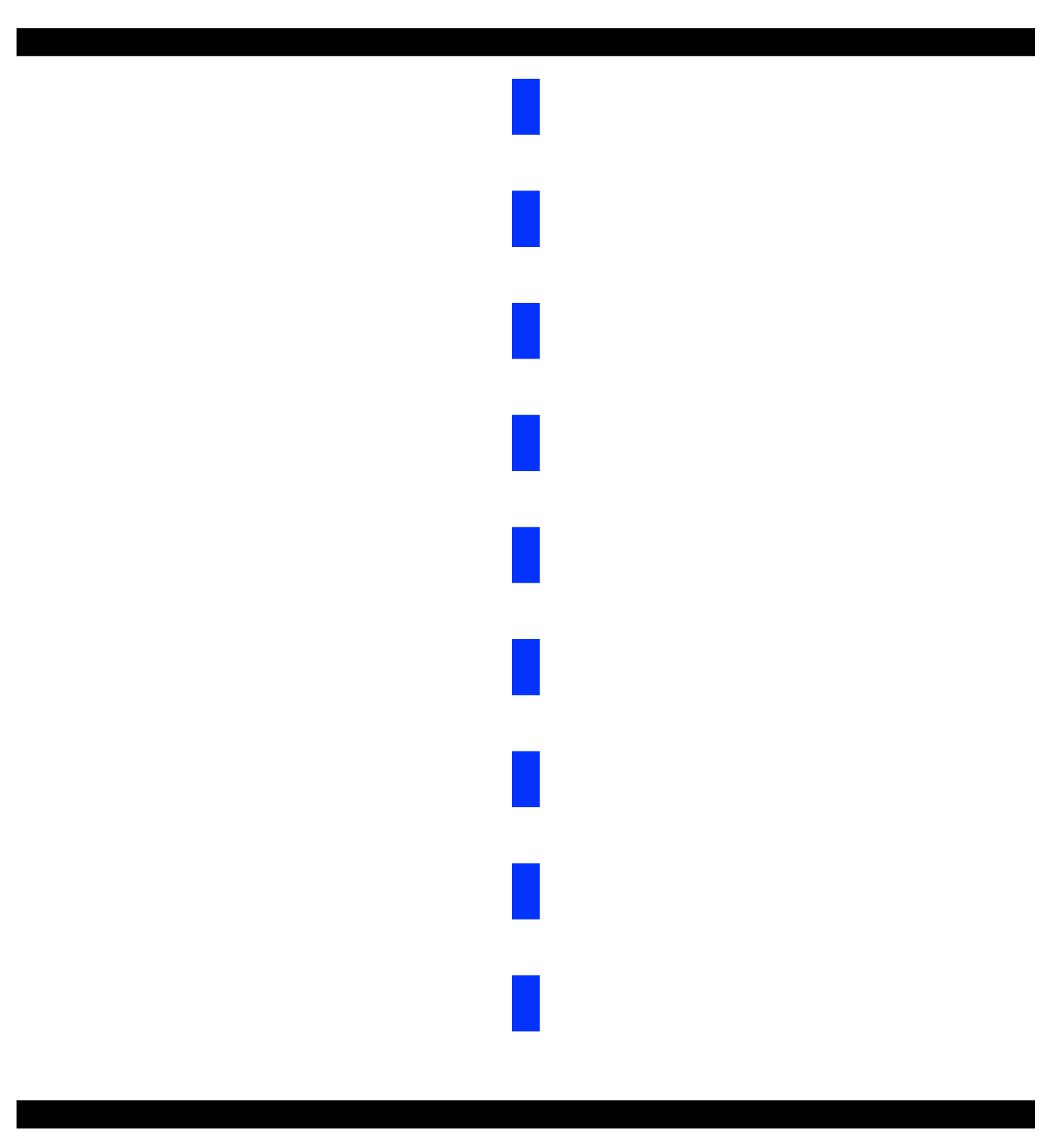}
\end{center}
\vspace*{-0.4cm}
\caption{The 0PN diagram is shown.\label{fig:NewtonGraph}}
\end{figure}

\noindent
In fig.~\ref{fig:vertices} the four vertices which are required for the
calculation of the static contribution to the two-body potential up to
5th PN order are shown.  From the point-particle action one obtains the
first vertex of fig.~\ref{fig:vertices}, while the remaining three
originate from the bulk action. The bulk vertices are distinguished by
the fact that they contain either zero or two scalar fields.
A typical Feynman diagram is then for example given by the tree-level 
graph which is shown in fig.~\ref{fig:NewtonGraph}. Its calculation delivers
the well known Newton potential. 
\section{Calculational strategy and factorization property\label{sec:calc}}
\begin{figure}
\begin{center}
\begin{minipage}{4cm}
  \includegraphics[width=3.6cm]{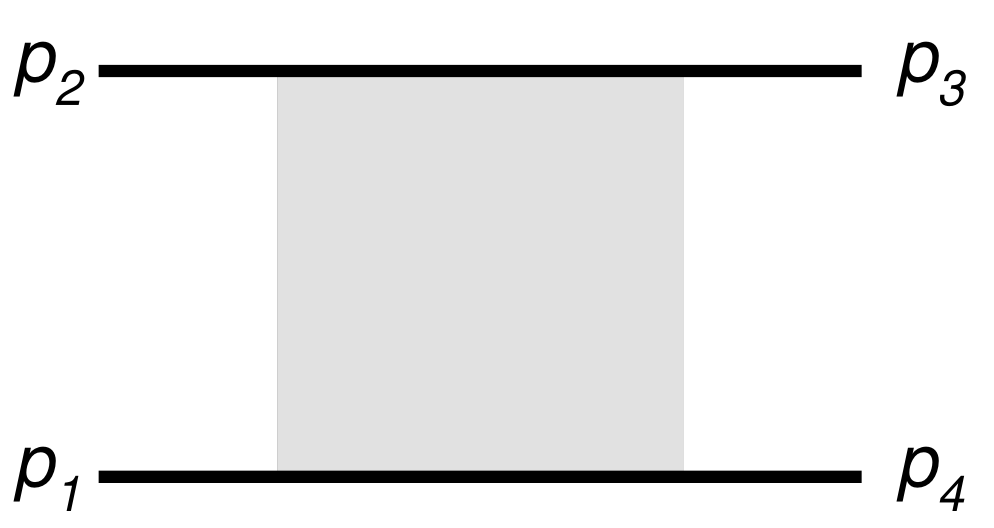} 
\end{minipage}
\hspace{1.4ex}=\hspace{2.8ex}
\begin{minipage}{4cm}
  \includegraphics[width=1.4cm]{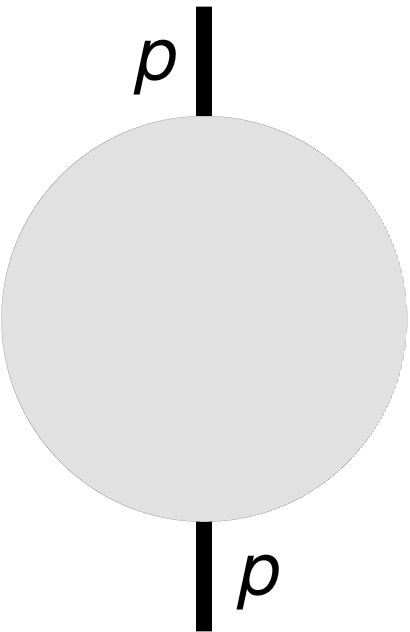} 
\end{minipage}
\end{center}
\vspace*{-0.4cm}
  \caption{An illustration of the mapping of gravity Feynman diagrams to
    self-energies as used in ref.~\cite{Foffa:2016rgu}.
  \label{fig:BoxSelf}} 
\end{figure}

For the calculation of the static part of the two-body potential up to
fifth PN order we consider only the classic contribution of our Feynman
diagrams and do not take into account any quantum corrections.
In general the original gravity Feynman diagrams depend on the ingoing
and outgoing momenta $p_1$, $p_2$, $p_3$ and $p_4$, like shown in
fig.~\ref{fig:BoxSelf}, however, it turns out that the corresponding
loop integrals are only functions of the momentum transfer
$p_3-p_2=p=p_1-p_4$.  As a result of this the corresponding loop
integrals can be represented by self-energy type diagrams as illustrated
in fig.~\ref{fig:BoxSelf}, see also ref.~\cite{Foffa:2016rgu}.
The contribution of a given amplitude to the two-body potential $V$ is
obtained by performing the Fourier transform, i.e. going from momentum
to coordinate space:
\begin{equation}
V=\ri \lim_{d\to3}\int_p e^{\ri p\cdot r}\;\;
\raisebox{-0.8cm}{\includegraphics[width=1.2cm]{QFT_bubble}}\;\;\;\sim\;\;\;
\raisebox{-0.5cm}{\includegraphics[width=1.3cm]{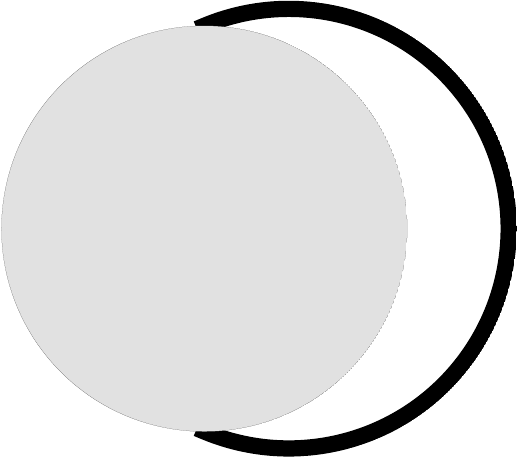}}\;\longrightarrow\;
\raisebox{-0.4cm}{\includegraphics[width=1.2cm]{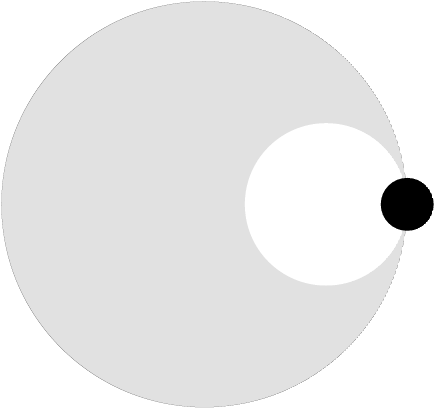}}\;.
\label{eq:Fourier}
\end{equation}
In ref.~\cite{Foffa:2019hrb} a theorem was shown, which states that
\begin{center}
  {\it{static graphs at odd $(2n+1)$-PN orders are factorizable}},
\end{center}
with $n\in\mathbb{N}_0$. A factorizable graph has at least one
matter-$\phi^k$ vertex with $k>1$ (see. fig.~\ref{fig:vertices}), while
a prime graph contains only matter-$\phi^k$ vertices with
$k=1$~\cite{Foffa:2019hrb}. The factorization property allows one to
recursively determine a given odd PN order from the known lower PN
ones. This strongly simplifies the determination of odd higher PN orders
compared to performing a direct calculation of the appearing loop
integrals, which becomes increasingly difficult with an increasing
number of loops. The factorization becomes apparent diagrammatically
when doing the Fourier transform of the amplitude to coordinate space as
shown in eq.~(\ref{eq:Fourier}). The additional integration over $p$ can
be interpreted as an additional loop integration, which can be
visualized by joining the external legs of the self-energy into an
additional propagator-like line and pinching it to a
point~\cite{Foffa:2019hrb} as it is illustrated on the r.h.s. of
eq.~(\ref{eq:Fourier}).

The contribution of a factorizable graph to the potential can then just
be obtained by multiplying together the results for the lower PN
subgraphs~\cite{Foffa:2019hrb}
\begin{equation}
\label{eq:factorization}
V_{n}^{\rm{\tiny{factorizable}}}=\left(V_{L,n_{1}}\times V_{R,n_{2}}\right)\times{\mathcal{K}}\times{\mathcal{C}},
\end{equation}
with $n=n_{1}+n_{2}+1$, where $V_{L,n_{1}}$ and $V_{R,n_{2}}$ is the
potential of the left and right subgraph. The factor $\mathcal{K}$ takes
into account the new matter-$\phi^k$ vertex which emerges through
glueing the two subgraphs together. The factor $\mathcal{C}$ is a
combinatoric factor.

In order to illustrate this method let us consider the well known static
odd 1PN and 3PN orders.
The 1PN potential can be obtained from a single static
diagram. According to the factorization theorem, it can be decomposed in
terms of two Newton diagrams
\begin{eqnarray}\label{eq:1PN}
\hspace*{0.7cm}
\begin{minipage}{2.0cm}
\begin{center}
\includegraphics[width=2.5cm]{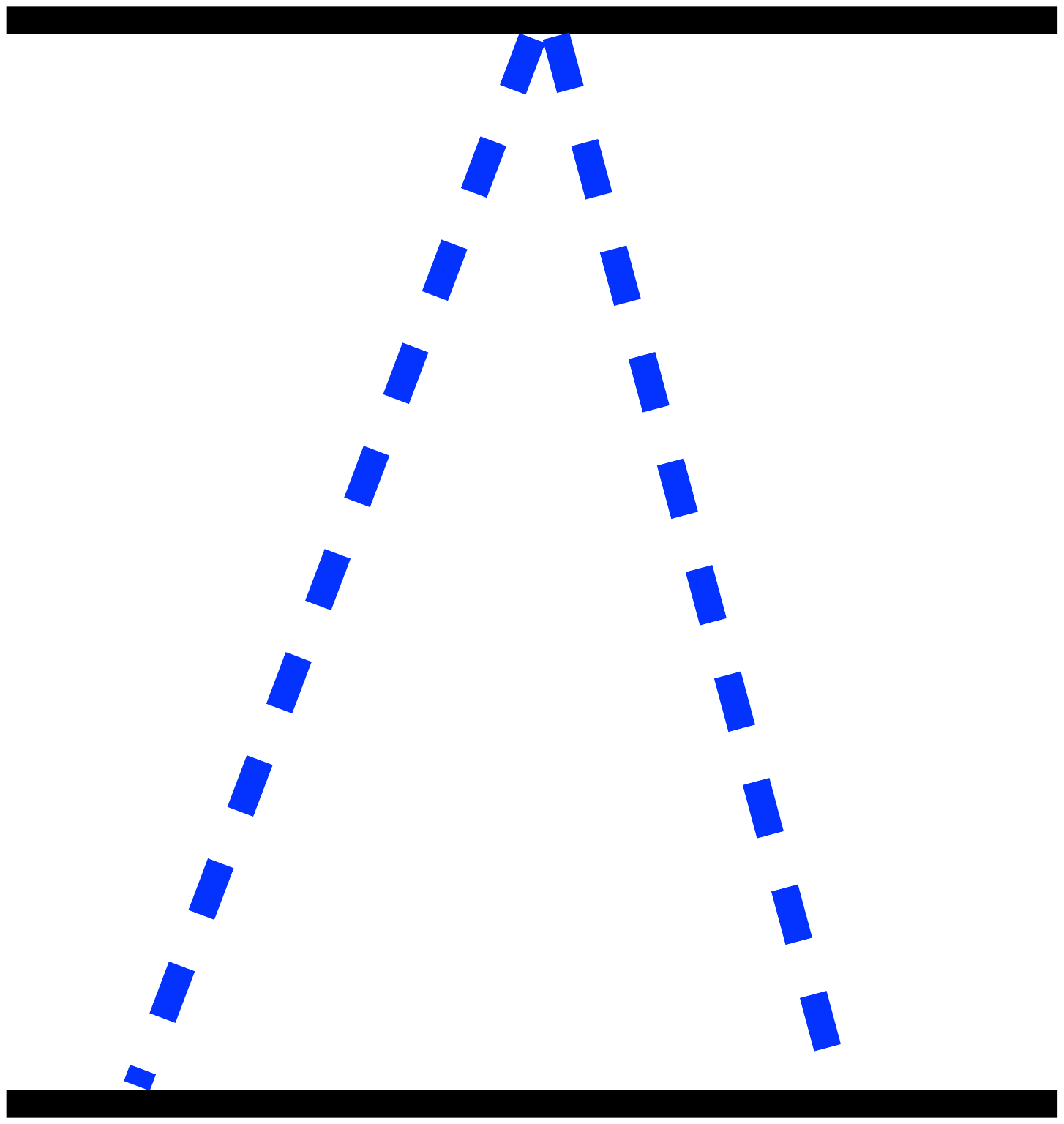}
\end{center}
\end{minipage}
=
\left(\begin{minipage}{1.5cm}
\begin{center}
\includegraphics[width=1.3cm]{Newton.pdf}
\end{center}
\end{minipage}\right)^2
\times
\frac{\begin{minipage}{1.2cm}
%\begin{center}
\hspace*{-0.25cm}
\includegraphics[width=1.3cm]{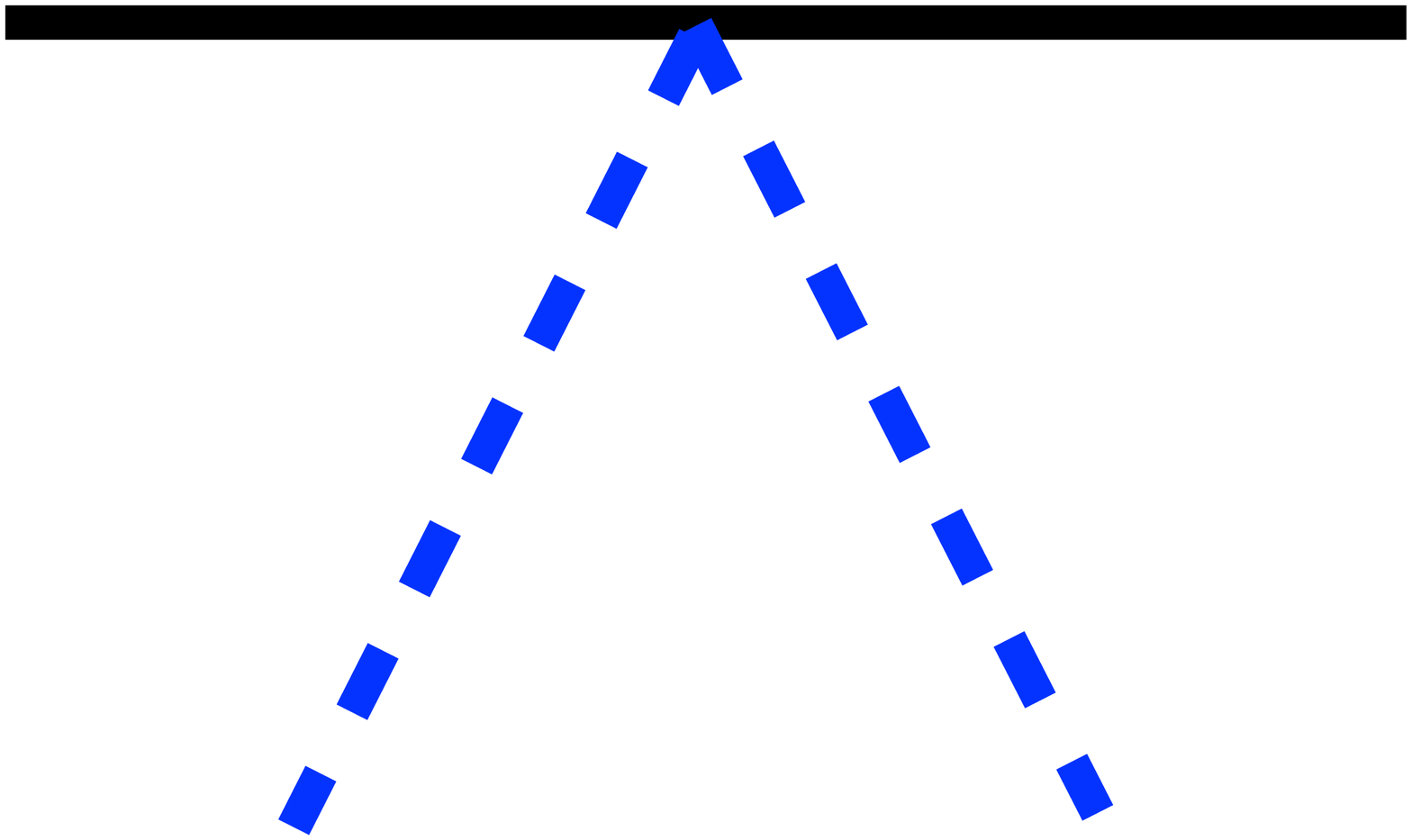}
%\end{center}
\end{minipage}}
{\left(\begin{minipage}{1.1cm}
%\begin{center}
\hspace*{-0.15cm}
\includegraphics[width=1.3cm]{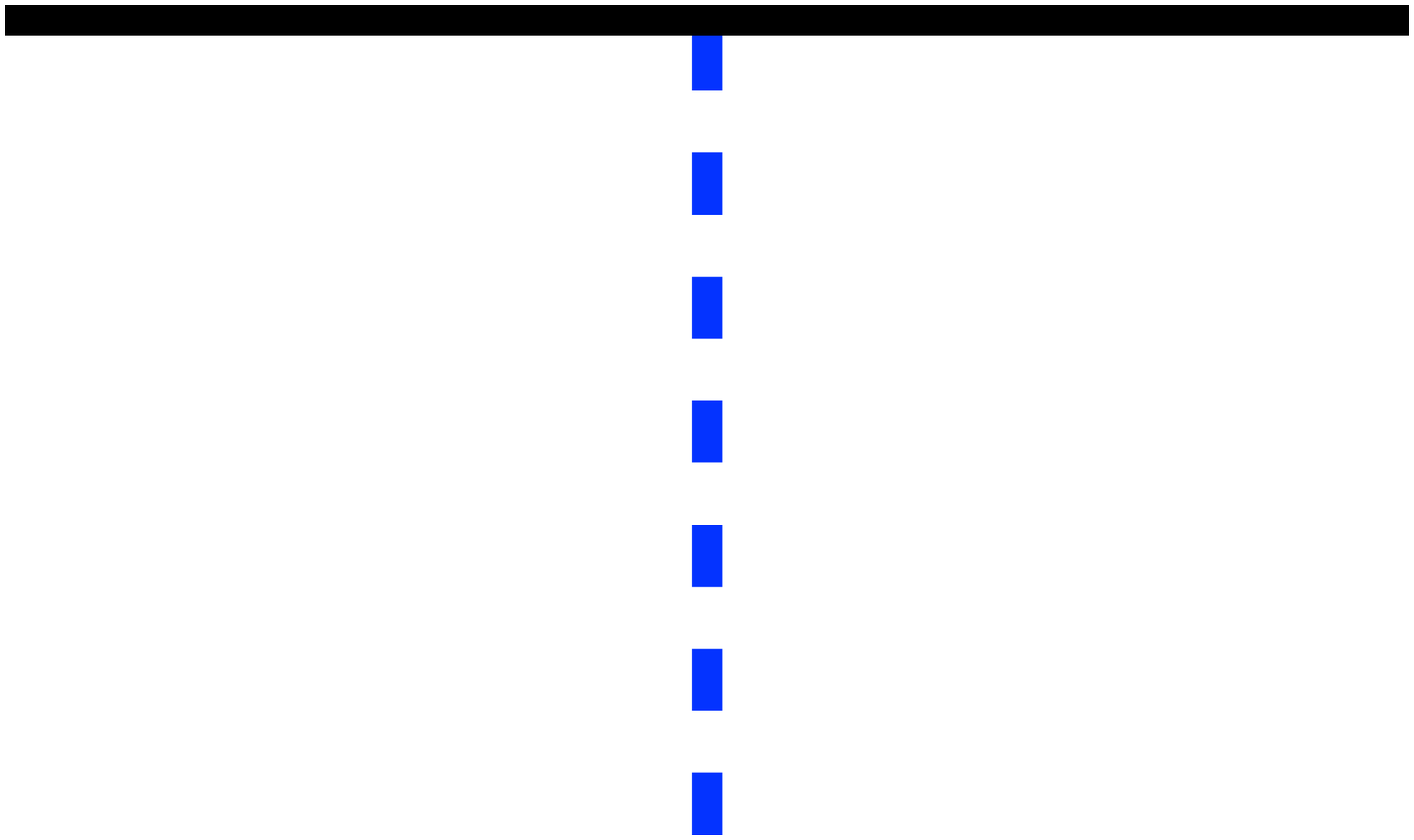}%
%\end{center}
\end{minipage}\right)^2}\,,
\end{eqnarray}
where the individual factors on the $r.h.s.$ indeed lead to the known result
for the one-loop diagram,
\begin{eqnarray}
\;\;\;\qquad%
\frac{G_N^2 m_1^2 m_2}{2r^2}=
\left(\!\!-\frac{G_N m_1  m_2}{r}\right)^2\!\times
{{G_N m_2/2} \over \left(\sqrt{G_N} m_2\!\right)^{2}}\,.
\end{eqnarray}
In this example, the factor ${\cal K}$ of eq.~(\ref{eq:factorization})
is given by the fraction on the $r.h.s.$ of eq.~(\ref{eq:1PN}), while
${\cal C}=1$. We have adopted the convention that $m_{1(2)}$ refers to
the bottom (top) line in the diagrams. The static 1PN potential is then
given by ${V}^{({\rm 1PN})}_{\rm static} =\frac{G_N^2 m_1^2
  m_2}{2r^2}+(m_1\leftrightarrow m_2)$.\\

\begin{figure}[h]
\begin{center}
\begin{minipage}{2.3cm}
\begin{center}
\includegraphics[width=2.3cm]{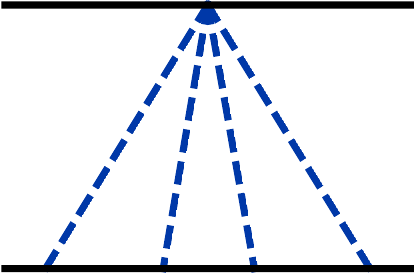}\\
$\mbox{\footnotesize{A}}_{\mbox{\tiny{3PN}}}$
\end{center}
\end{minipage}
\hspace*{0.4cm}
\begin{minipage}{2.3cm}
\begin{center}
\includegraphics[width=2.3cm]{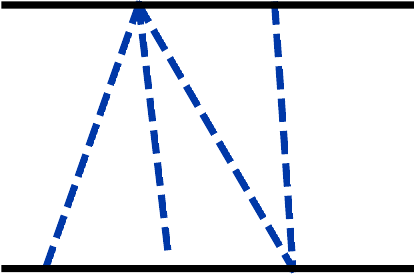}\\
$\mbox{\footnotesize{B}}_{\mbox{\tiny{3PN}}}$
\end{center}
\end{minipage}
\hspace*{0.4cm}
\begin{minipage}{2.3cm}
\begin{center}
\includegraphics[width=2.3cm]{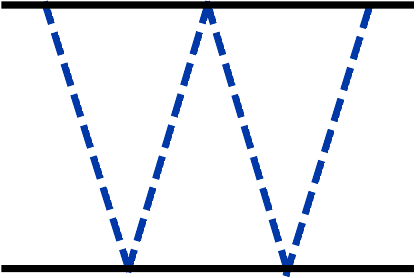}\\
$\mbox{\footnotesize{C}}_{\mbox{\tiny{3PN}}}$
\end{center}
\end{minipage}
\hspace*{0.4cm}
\begin{minipage}{2.3cm}
\begin{center}
\includegraphics[width=2.3cm]{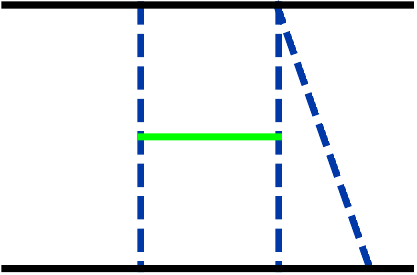}\\
$\mbox{\footnotesize{D}}_{\mbox{\tiny{3PN}}}$
\end{center}
\end{minipage}\\[0.5cm]
\hspace*{0.0cm}
\begin{minipage}{2.3cm}
\begin{center}
\includegraphics[width=2.3cm]{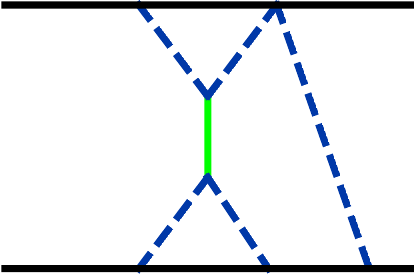}\\
$\mbox{\footnotesize{E}}_{\mbox{\tiny{3PN}}}$
\end{center}
\end{minipage}
\hspace*{0.4cm}
\begin{minipage}{2.3cm}
\begin{center}
\includegraphics[width=2.3cm]{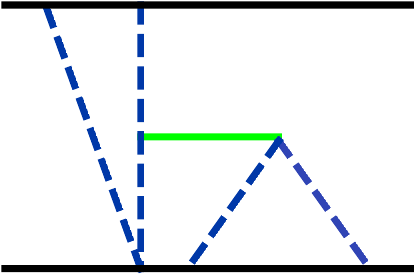}\\
$\mbox{\footnotesize{F}}_{\mbox{\tiny{3PN}}}$
\end{center}
\end{minipage}
\hspace*{0.4cm}
\begin{minipage}{2.3cm}
\begin{center}
\includegraphics[width=2.3cm]{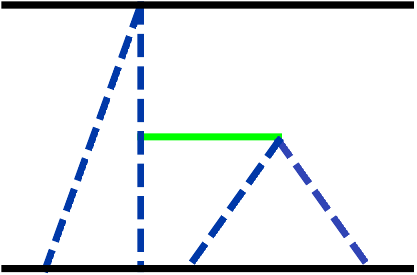}\\
$\mbox{\footnotesize{G}}_{\mbox{\tiny{3PN}}}$
\end{center}
\end{minipage}
\hspace*{0.4cm}
\begin{minipage}{2.3cm}
\begin{center}
\includegraphics[width=2.3cm]{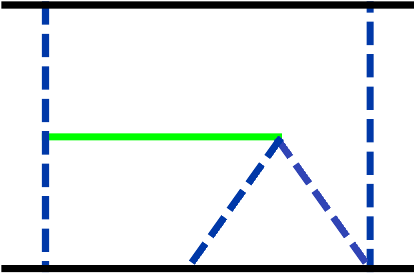}\\
$\mbox{\footnotesize{H}}_{\mbox{\tiny{3PN}}}$
\end{center}
\end{minipage}
\caption{The 3PN graphs are shown. The first graph in the second line
  (bottom-left) does not contribute to the 3PN potential, because its
  2PN subdiagram vanishes.} 
\label{fig:3PNdiagrams}
\end{center}
\end{figure}

At the 3PN order there are eight static graphs which are shown in
fig.~\ref{fig:3PNdiagrams}.
Their contributions to the Lagrangian were computed in
\cite{Foffa:2011ub}.  In light of the factorization theorem, we identify
two classes of diagrams:

\noindent
{\textbf{1.}} In this set, we consider three diagrams composed out of four Newtonian
factors, corresponding to the first three diagrams of
fig.~\ref{fig:3PNdiagrams}
($\mbox{A}_{\mbox{\footnotesize{3PN}}}$,
$\mbox{B}_{\mbox{\footnotesize{3PN}}}$,
$\mbox{C}_{\mbox{\footnotesize{3PN}}}$). 
We represent them as:
\begin{eqnarray}
\left(\begin{minipage}{1.7cm}
%\begin{center}
\hspace*{0.1cm}
  \includegraphics[width=1.3cm]{Newton.pdf}
%\hspace*{0.1cm}
%\end{center}
\end{minipage} \right)^4\,.
\end{eqnarray}
Their contribution to the 3PN potential is:
\begin{eqnarray}
{V}_{N^4} &=&
\frac{1}{24}\frac{G_N^4 m_1^4 m_2}{r^4}+\frac{G_N^4 m_1^3
                    m_2^2}{r^4}+(m_1\leftrightarrow m_2) \ .
\qquad
\end{eqnarray}

\noindent
{\textbf{2.}}
The other five diagrams of fig.~\ref{fig:3PNdiagrams}
($\mbox{D}_{\mbox{\footnotesize{3PN}}}$,
$\mbox{E}_{\mbox{\footnotesize{3PN}}}$,
$\mbox{F}_{\mbox{\footnotesize{3PN}}}$,
$\mbox{G}_{\mbox{\footnotesize{3PN}}}$,
$\mbox{H}_{\mbox{\footnotesize{3PN}}}$), are built as
products of one Newtonian term and the three
static 2PN prime graphs, combined in all possible ways,
schematically represented as:
\begin{eqnarray}
\begin{minipage}{1.2cm}
\begin{center}
\includegraphics[width=1.3cm]{Newton.pdf}
\end{center}
\end{minipage}
\!\!\!\times
\left(\begin{minipage}{3.6cm}
\begin{center}
\includegraphics[width=3.4cm]{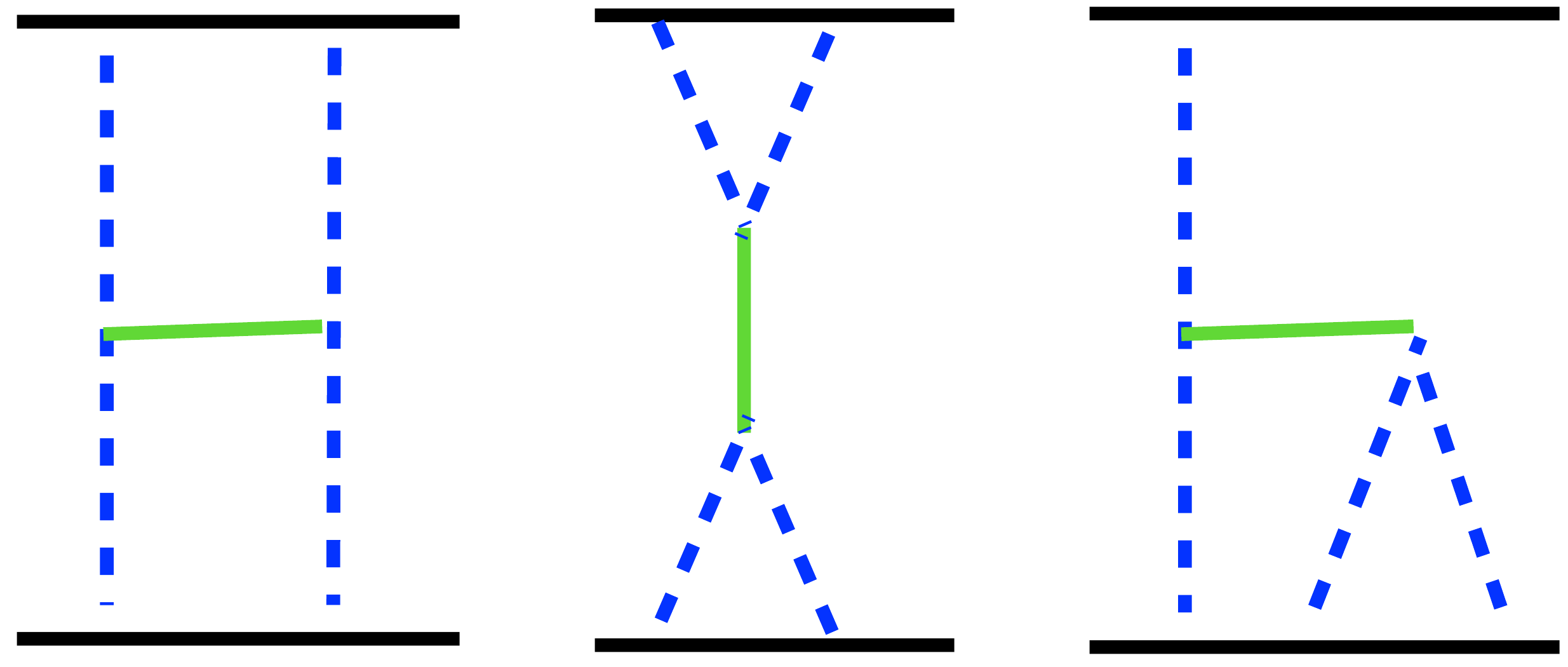}
\end{center}
\end{minipage}\right) \ .
\end{eqnarray}
For illustration purposes, the factorization theorem can be verified for one of them:
\begin{eqnarray}
\qquad
\begin{minipage}{2.7cm}
\begin{center}
\includegraphics[width=2.7cm]{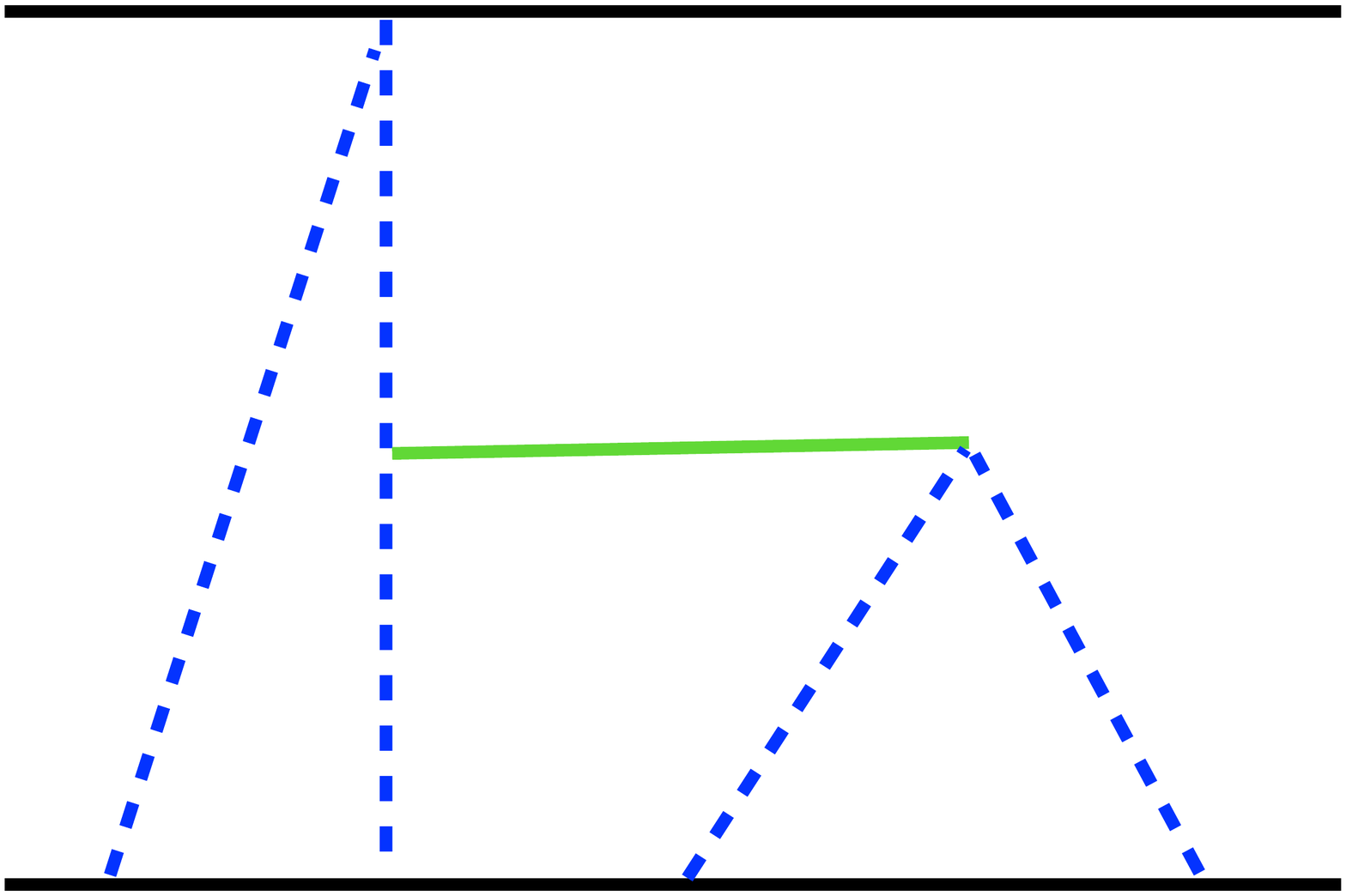}
\end{center}
\end{minipage}
\hspace*{0ex}=\hspace*{1ex}
2 \times\hspace*{1ex} \begin{minipage}{1.5cm}
\begin{center}
\includegraphics[width=1.3cm]{Newton.pdf}
\end{center}
\end{minipage}
\hspace*{1ex}\times\hspace*{0ex}
\begin{minipage}{2.45cm}
\begin{center}
\includegraphics[width=2.45cm]{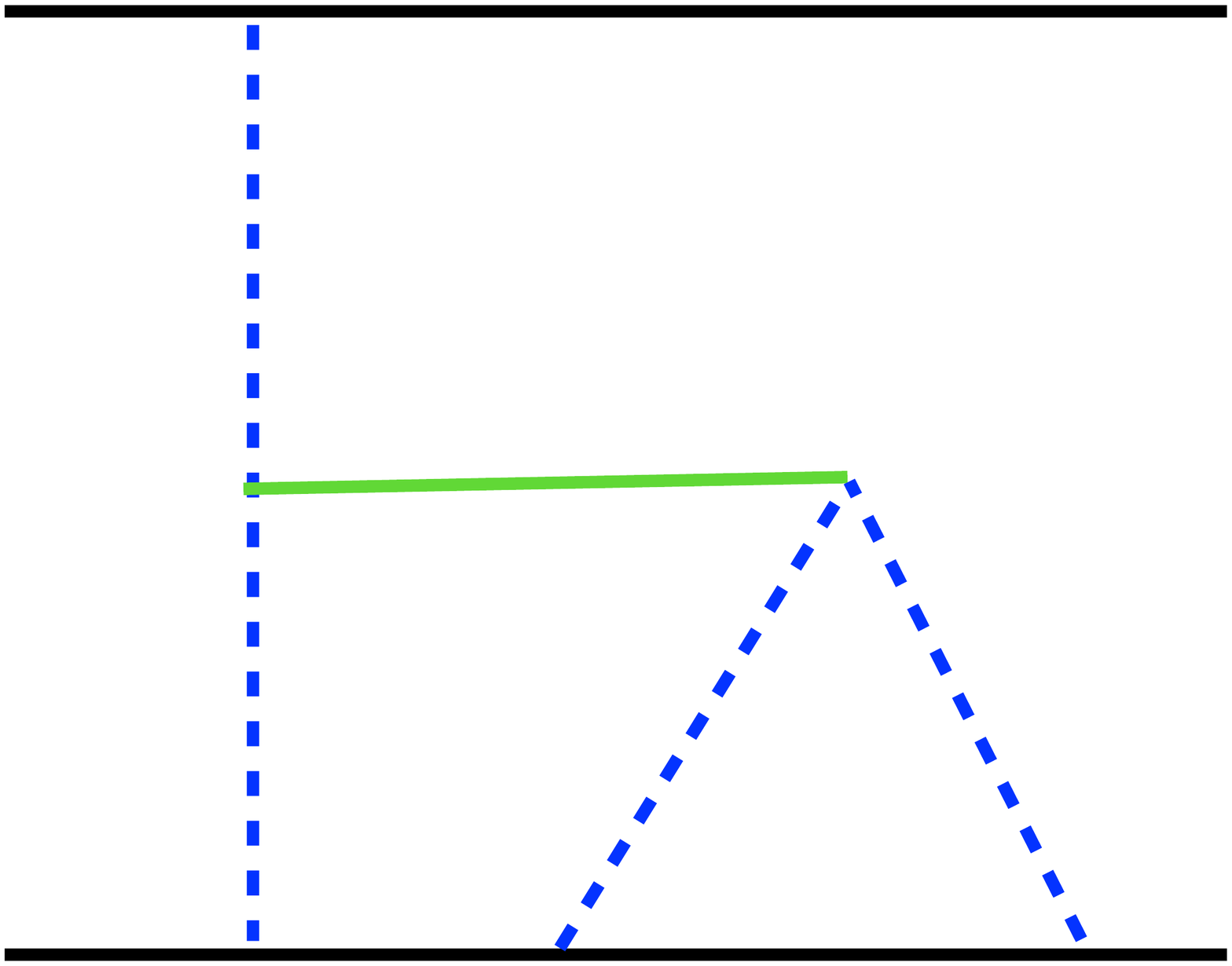}
\end{center}
\end{minipage}
\!\!\!\times\;\;
\frac{\begin{minipage}{1.2cm}
\begin{center}
\hspace*{-0.25cm}
\includegraphics[width=1.3cm]{particle2.pdf}
\end{center}
\end{minipage}}
{\left(\begin{minipage}{1.1cm}
\begin{center}
\hspace*{-0.15cm}
\includegraphics[width=1.3cm]{particle1.pdf}
\end{center}
\end{minipage}\right)^2}\,,
\qquad\quad
\end{eqnarray}
(${\cal C}=2$ here) amounting to
\begin{equation}
\quad\qquad\frac{G_N^4 m_1^4 m_2}{3r^4}\;\;=\;\;
2
 \times 
\left(\!-\frac{G_N m_1 m_2}{r} \right) 
 \times 
\left(\!-\frac{G^3_N m_1^3 m_2}{3r^3} \right) 
 \times 
{ {G_N m_2/2} \over \left(\sqrt{G_N} m_2\right)^{2}}\,,\qquad 
\end{equation} 
in agreement with \cite{Foffa:2011ub}.
The contribution to the potential from the five diagrams in class 2 is: 
\begin{eqnarray}
{V}_{N \times {\rm 2PN}} &=&
\frac{1}{3}\frac{G_N^4 m_1^4 m_2}{r^4}+5\frac{G_N^4 m_1^3
                    m_2^2}{r^4}+(m_1\leftrightarrow m_2) \, .
\qquad 
\end{eqnarray}
The total static 3PN contribution of the diagrams belonging to
classes 1 and 2 is, in agreement with the literature:
\begin{equation}
{V}^{({\rm 3PN})}_{\rm static} = 
{V}_{N^4} + {V}_{N \times {\rm 2PN}}
= \frac{3}{8}\frac{G_N^4 m_1^4 m_2}{r^4}+6\frac{G_N^4 m_1^3
                    m_2^2}{r^4}+(m_1\leftrightarrow m_2)\,.
\end{equation}

\section{The results\label{sec:results}}
The static contribution to the two-body potential at fifths PN order can
be determined recursively from the lower PN results with the help
of the factorization property of ref.~\cite{Foffa:2019hrb} as discussed
in sec.~\ref{sec:calc}. The 154 five-loop diagrams can be subdivided
here into four classes according to their factorization property.  The
first class consist out of products of six Newtonian diagrams. The
second class consists out of products of three Newtonian graphs and
the prime 2PN diagrams. The third class are products of a Newtonian
diagram and the 4PN prime graphs. Finally the last class are products
of the 2PN prime graphs.

\begin{figure}
\begin{center}
\begin{minipage}{3cm}
\begin{center}
\includegraphics[width=2.1cm]{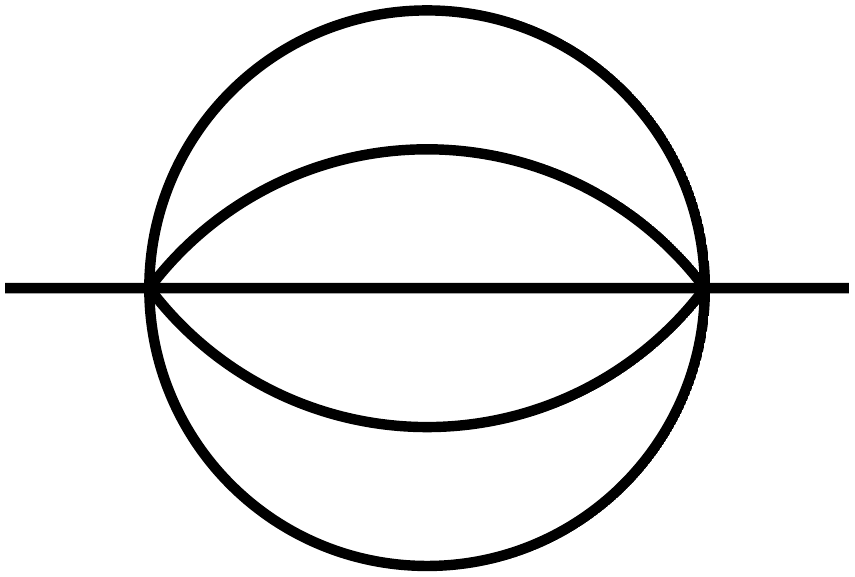}\\
$\mathcal{M}_{0,1}$
%1040501\\
%$\vep^2$
\end{center}
\end{minipage}
\begin{minipage}{3cm}
\begin{center}
\includegraphics[width=2.1cm]{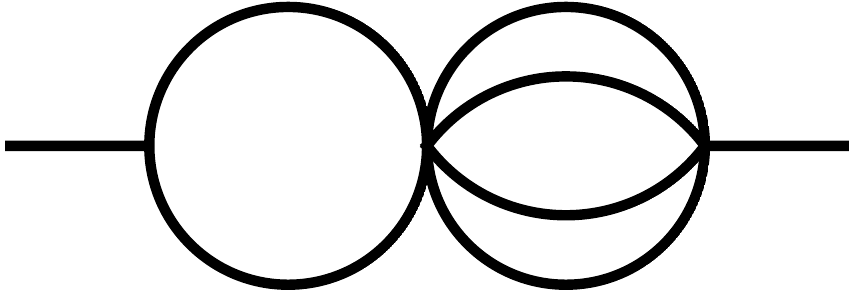}\\
$\mathcal{M}_{1,1}$
%1040605\\
%$\vep^0$
\end{center}
\end{minipage}
\begin{minipage}{3cm}
\begin{center}
\includegraphics[width=2.1cm]{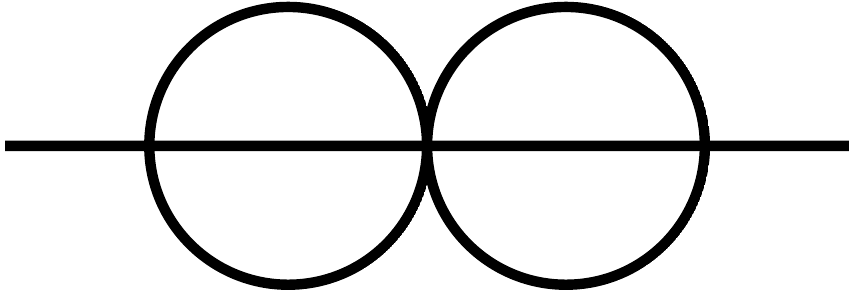}\\
$\mathcal{M}_{1,2}$
%1040607\\
%$\vep^1$
\end{center}
\end{minipage}\\[0.3cm]
%%%%%%%%%%
%%
\begin{minipage}{3cm}
\begin{center}
\includegraphics[width=2.1cm]{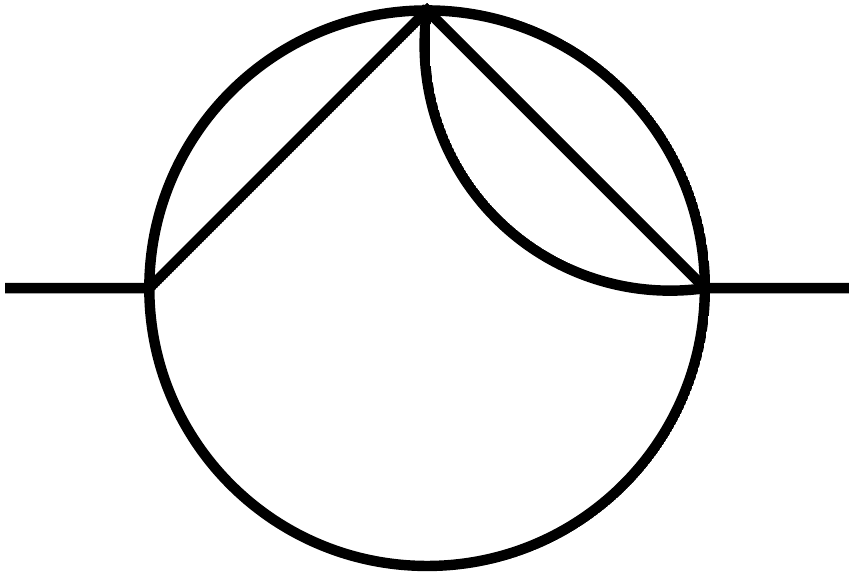}\\
$\mathcal{M}_{1,3}$
\end{center}
\end{minipage}
\begin{minipage}{3cm}
\begin{center}
\includegraphics[width=2.1cm]{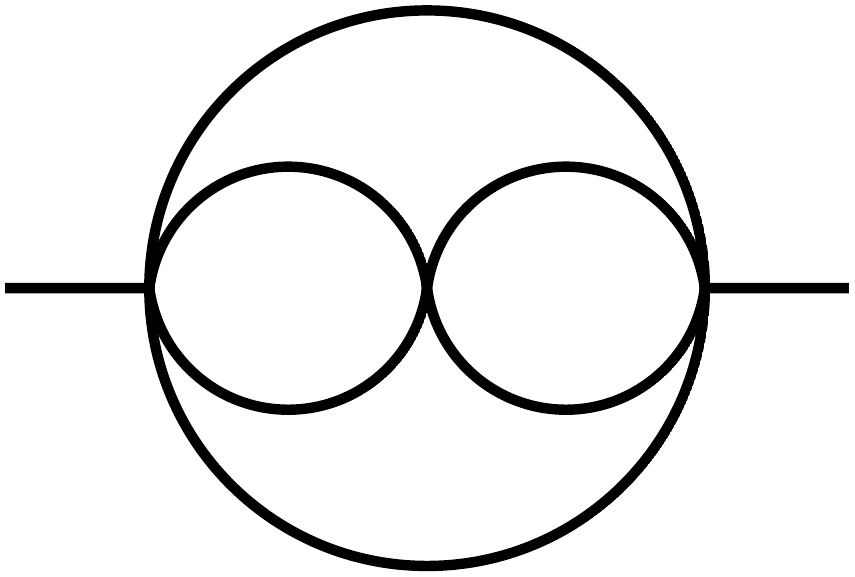}\\
$\mathcal{M}_{1,4}$
\end{center}
\end{minipage}
\begin{minipage}{3cm}
\begin{center}
\includegraphics[width=2.1cm]{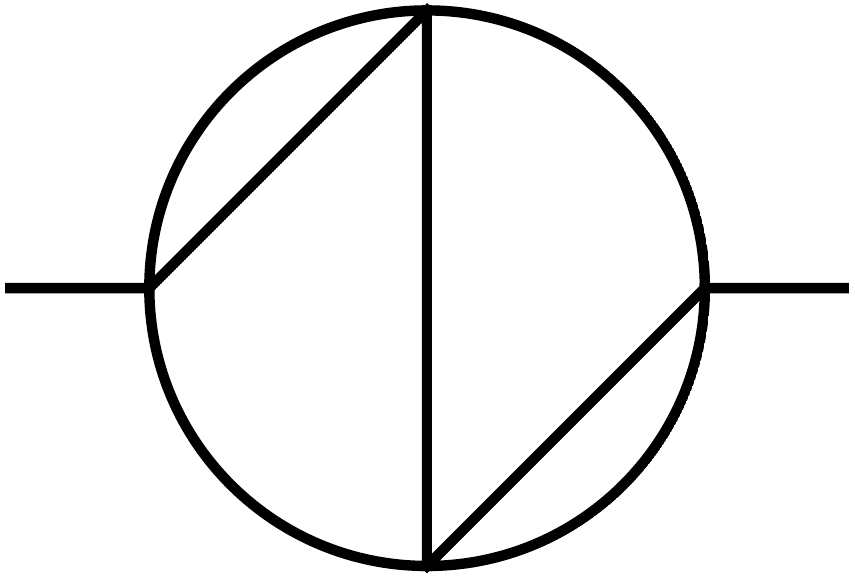}\\
$\mathcal{M}_{2,2}$
\end{center}
\end{minipage}
\begin{minipage}{3cm}
\begin{center}
\includegraphics[width=2.1cm]{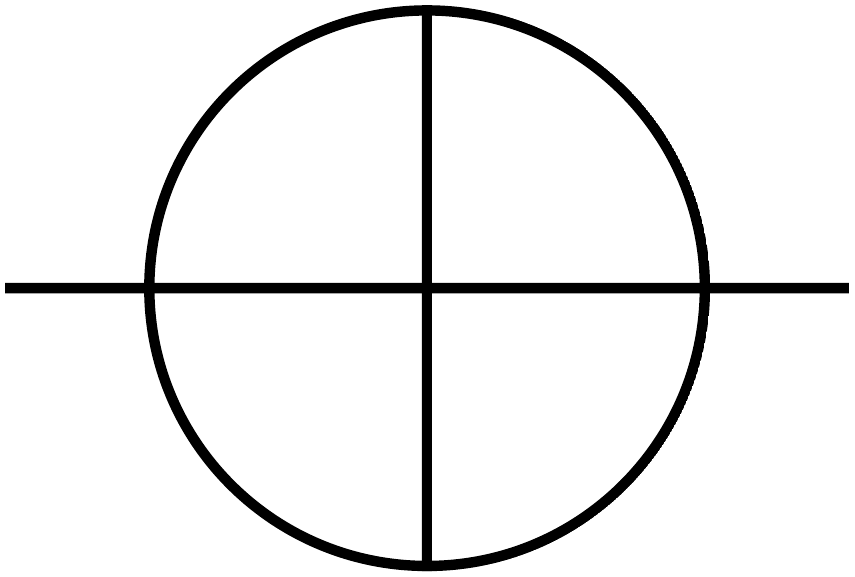}\\
$\mathcal{M}_{3,6}$
\end{center}
\end{minipage}
\end{center}
\caption{The master integrals appearing in the calculation of the static
  4PN contribution are shown.\label{fig:MI}}
\end{figure}

The ingredients which are needed in order to determine the static 5PN
contribution are only the static 2PN and 4PN prime graphs. All static 5PN
contributions can then be obtained through the factorization
property. The static 2PN and 4PN prime graphs are known since long from
literature, as discussed in sec.~\ref{sec:Intro}. In particular the
static 4PN contribution has been computed in the EFT approach in
ref.~\cite{Foffa:2016rgu} by employing techniques for calculating
Feynman diagrams which are used in high energy particle physics. The
appearing 50 diagrams are subdivided into two sets. The first set
contains simpler integrals which can be computed with the kite
rule~\cite{Tkachov:1981wb,Chetyrkin:1981qh}. The second set of four-loop
integrals has been reduced systematically to a small set of seven
master integrals~(MI) with integration-by-parts
identities~\cite{Tkachov:1981wb,Chetyrkin:1981qh} using Laporta's
algorithm~\cite{Laporta:1996mq,Laporta:2001dd}. They are shown in
fig.~\ref{fig:MI}.
The reduction has been performed in two ways, with an in-house code
based on
{\texttt{Form}}~\cite{Vermaseren:2000nd,Vermaseren:2002rp,Tentyukov:2006ys}
and with the program
{\texttt{Reduze}}~\cite{Studerus:2009ye,vonManteuffel:2012np}.  Five of
the seven MIs $\{\mathcal{M}_{0,1}, \mathcal{M}_{1,1},
\mathcal{M}_{1,2}, \mathcal{M}_{1,3}, \mathcal{M}_{1,4}\}$ can be
calculated straightforwardly in a closed analytical form in $d$
dimensions. They are expressible in terms of $\Gamma$-functions. The MI
$ \mathcal{M}_{2,2}$ always appears multiplied by sufficient high
positive powers of $\varepsilon=(d-3)$ in the amplitude, so that it
drops out in the limit $\varepsilon\to0$.  The remaining seventh MI
$\mathcal{M}_{3,6}$ is known in an expansion in $\varepsilon$
in refs.~\cite{Lee:2015eva,Foffa:2016rgu,Damour:2017ced}.

Having all even PN orders at hand, the static 5PN contribution to the
gravitational two-body potential has been obtained for the first time in
ref.~\cite{Foffa:2019hrb} with the help of the factorization
property. It reads:
\begin{equation}
\label{eq:V5PN}
{V}^{({\rm 5PN})}_{\rm static} =
 {5\over16}{G_N^6m_1^6m_2\over r^6}
+{91\over6}{G_N^6m_1^5m_2^2\over r^6}
+{653\over6}{G_N^6m_1^4m_2^3\over r^6}
+(m_1\leftrightarrow m_2).
\end{equation}
The result has been checked in the test-particle limit, in which one
considers the body with mass $m_2$ as a test particle in the
gravitational field of the body with mass $m_1$, corresponding
to the Schwarzschild limit which recovers the term linear in $m_2$
and with the highest power in $m_1$. 
At fifth PN order this check has been used for the first-time in
ref.~\cite{Foffa:2019hrb}. 
In this limit the static effective Lagrangian reads $\mathcal{L}^{m_1\gg m_2}_{\rm static}
=-m_2\sqrt{1-G_Nm_1/r}/\sqrt{1+G_Nm_1/r}$. 
Its expansion permits to extract this contribution to the potential at each PN order. 
Hence, the Schwarzschild metric allows to predict the coefficients
of the terms of the form ${G_N^{\ell}m_1^{\ell}m_2 / r^\ell}$ at any
$n$-th PN order with $\ell=n+1$, for example, at 6th PN order the
coefficient of the term ${G_N^7m_1^7m_2 / r^7}$ reads $-5/16$. 
Finally eq.~(\ref{eq:V5PN}) has been confirmed in ref.~\cite{Blumlein:2019zku}
by an independent calculation.

\section{Summary and conclusion\label{sec:conclude}}
We studied the gravitational two-body potential at fourth and fifth PN
order in the EFT approach to GR in the static limit. Its calculation can
be mapped onto the determination of four- and five-loop self-energies,
which can be solved with tools commonly used in high-energy particle
physics. We established a factorization property of the static diagrams
appearing at odd PN orders, so that these contributions can be
determined recursively from the lower PN order results and no loop
integrals need to be computed. We verified the validity of our
factorization theorem at the lower odd PN orders and applied it to the
fifth PN order in order to do a first-time calculation of the static
contributions to the gravitational two-body potential. The factorization
property is also applicable to a large subset of even-PN diagrams, which
simplifies their calculation. As a result of this the factorization
property is a powerful tool to simplify higher order PN calculations.

\acknowledgments
S.F. has been supported by the Fonds National Suisse and by the SwissMap
NCCR.
P.M. has been supported by the Supporting TAlent in ReSearch at Padova
University (UniPD STARS Grant 2017 "Diagrammalgebra").
RS is partially supported by CNPq.
W.J.T. has been supported in part by Grants No. FPA2017-84445-P and
No. SEV-2014-0398 (AEI/ERDF, EU), the COST Action CA16201 PARTICLEFACE,
and the "Juan de la Cierva Formaci\'on" program (FJCI-2017-32128).
%

%%
%%Literature
%%
\providecommand{\href}[2]{#2}\begingroup\raggedright\endgroup
\end{document}